\documentclass[article]{elsart}
\usepackage{epsfig}

\begin{document}

\begin{frontmatter}

\title{Saturation properties of nuclear matter in a relativistic mean field model constrained by the quark dynamics}

\author{R. Huguet, J.C. Caillon and J. Labarsouque}

\address{Centre d'Etudes Nucl\'{e}aires de Bordeaux-Gradignan, CNRS-IN2P3 \\ Universit\'{e} Bordeaux 1, Le Haut-Vigneau, 33170 Gradignan Cedex, France}

\begin{abstract}
We have built an effective Walecka-type hadronic Lagrangian in which the hadron masses and
the density dependence of the coupling constants are deduced from the quark dynamics using a Nambu-Jona-Lasinio model. In order to stabilize nuclear matter an eight-quark term has been included. The parameters of this Nambu-Jona-Lasinio model have been determined using the
meson properties in the vacuum but also in the medium through the omega
meson mass in nuclei measured by the TAPS collaboration. Realistic properties of nuclear matter have been obtained. 
\end{abstract}



\end{frontmatter}

\section{Introduction}
One of the most fascinating challenges of nuclear physics is the description of nuclear matter and nuclei starting from Quantum ChromoDynamics (QCD). Even if important progress have been made in QCD calculations on the lattice, such a description is not yet available. Models incorporating the most prominent features of QCD have to be used.

Quark models, like, for example, the quark-meson
coupling model\cite{gui,sai} which describes nuclear matter as non
overlapping nucleon MIT bags interacting through the self-consistent
exchange of meson fields directly coupled to the quarks, or the model of Bentz and Thomas\cite{b.t} where the nucleons are generated as 
quark-diquark states in a Nambu-Jona-Lasinio (NJL) model with an infrared cutoff to simulate the confinement, are gaining more and more success in shedding some light on the dynamical behaviour of nucleons in nuclei. 

On the other hand, starting directly at the hadronic level, Kohn-Sham-like density functional theories with constraints on the energy functional coming from low energy QCD, like that developed by Finelli, Kaiser, Vretenar and Weise\cite{finelli}, lead to an accurate description of nuclear matter and finite nuclei.

Another possibility is to apply the strategy of effective field theories where low energy effective hadronic Lagrangians are obtained by integrating out the degrees of freedom lying above the energy scale considered. This decimation leads to density-dependent masses and couplings in the hadronic Lagrangian \cite{br2}. This density-dependence should reflect essentially the underlying dynamics of the quarks. Such a Lagrangian with density dependent masses and coupling constants determined according to Brown and Rho scaling\cite{b.r} has been proposed by Brown, Song, Min and Rho\cite{son}. The calculation reported in \cite{son}, assuming a scaling law  leading to a decreasing of the vector meson mass of approximately 20\% at saturation, enables a realistic description of bulk properties of nuclear matter.

Recently, new experimental results for the in-medium $\omega$ meson mass would suggest a small decreasing of approximately 10-15\% at saturation density. In particular, the TAPS collaboration \cite{tap} found an in-medium mass of $m^*_{\omega} = 722^{+4}_{-4}\mathrm{(stat)}^{+35}_{-5}\mathrm{(syst)}$ MeV at 0.6 times the saturation density (which is compatible with a 14\% dropping at saturation when a linear density-dependence is assumed) in photoproduction experiments and Naruki et al. \cite{kek} found a 9\% decrease of the in-medium $\omega$ mass at saturation in 12 GeV proton-nucleus reactions.

In this paper, we explore the possibility of obtaining a realistic description of bulk properties of nuclear matter in the model of Brown, Song, Min and Rho \cite{son} with a density dependence of the in-medium $\omega$ mass in accordance with recent experimental indications.  In addition, in order to allow deviations from the Brown and Rho scaling, the hadron masses and the density dependence of the coupling constants have been deduced directly from the quark dynamics using an in-medium NJL model. We have chosen here to use the NJL model \cite{njl} since, despite some shortcomings like the lack of confinement, with its dynamical quark mass generation and the in-medium chiral symmetry restoration, it allows to take into account an important part of quark dynamics.  

The paper is organised as follows. In section 2, we recall the formalism of the effective hadronic model with density-dependent couplings and masses used to determine bulk properties of infinite nuclear matter. The section 3 will be devoted to the determination of density dependences for hadron masses and meson couplings from a NJL model.  We present and discuss numerical results in section 4, where some nuclear matter bulk properties (saturation point, incompressibility, nucleon effective mass ...) are presented. We conclude in section 5.

\section{Effective hadronic mean-field model with density-dependent parameters}

We use an effective hadronic Lagrangian similar to that used by Brown, Song, Min and Rho \cite{son}. It is formally identical to the Walecka one\cite{s.w} but with density-dependent masses and couplings, which reflect the underlying physics induced by the quark dynamics:

\begin{eqnarray}
\mathcal{L}_{had} &=&\overline{\psi }\left[ \gamma _{\mu }(i\partial ^{\mu
}-g_{\omega NN}\alpha_{\omega }^{*}V^{\mu })-(M_{N}^{*}-g_{\sigma NN}\alpha_{\sigma
}^{*}\phi )\right] \psi  \nonumber \\
&&+\frac{1}{2}(\partial _{\mu }\phi \partial ^{\mu }\phi -m_{\sigma
}^{*2}\phi ^{2})+\frac{1}{2}m_{\omega }^{*2}V_{\mu }V^{\mu }-\frac{1}{4}%
F_{\mu \nu }F^{\mu \nu },  \label{lag}
\end{eqnarray}

\noindent where $\psi $, $\phi $ and $V^{\mu }$ represent respectively the
nucleon, $\sigma $ and $\omega $ meson fields and, as usual, $F_{\mu \nu
}=\partial _{\mu }V_{\nu }-\partial _{\nu }V_{\mu }$. This Lagrangian contains only two free-parameters, $g_{\sigma NN}$ and $g_{\omega NN}$, which will be adjusted at the end of the calculation to reproduce the position of the saturation point.  

As already indicated, the density-dependent $\alpha^*_i$ functions and mass parameters, $M_{N}^{*}$, $m_{\sigma }^{*}$, $m_{\omega }^{*}$, would contain information coming from the quark level. The problem is now to determine these quantities in such a manner they reflect the underlying quark dynamics as better as possible. If a fully satisfactory description of nucleons and nuclear matter from quark and gluon fields would be available, this stage of calculation would be unnecessary since the fields and parameters entering $\mathcal{L}_{had}$ could be expressed directly from the quark level ones. However, it is not actually the case and approximations have still to be made. For example, in their work, Brown, Song, Min and Rho \cite{son} have presupposed a density dependence of the Lagrangian parameters in accordance with Brown and Rho scaling. In the present exploratory work, we have chosen to assume for the values of the masses and couplings the same baryonic density dependence as that which would be obtained in quark matter for the same density. As explained in the next section, we have used for this job a NJL model. Obviously, in the real world, quark matter cannot exist at low temperatures and densities but we expect that, despite some shortcomings, the approximation used is not so bad since we have verified that, for example, the density dependence of the quark condensate obtained is very similar to that obtained in ref\cite{b.t} using a quark-diquark description of the nucleons in nuclear matter.  

In the rest frame of symmetric nuclear matter, the equation for the nucleon
field in mean-field approximation is: 
\begin{equation}
\left[ i\gamma _{\mu }\partial ^{\mu }-g_{\omega NN}\alpha_{\omega }^{*}\gamma
^{0}V_{0}+\gamma ^{0}\Sigma _{0}-(M_{N}^{*}-g_{\sigma NN}\alpha_{\sigma }^{*}\phi
_{0})\right] \psi =0,  \label{eqn}
\end{equation}
where the scalar and vector classical fields $\phi _{0}$ and $V_{0}$ can be
obtained from the Lagrange equations :

\begin{equation}
\phi _{0}=\frac{g_{\sigma NN}\alpha_{\sigma }^{*}\rho _{s}}{m_{\sigma }^{*2}},
\label{fi0}
\end{equation}
\begin{equation}
V_{_{0\mu }}=\delta _{\mu 0}V_{0}=\delta _{\mu 0}\frac{g_{\omega
NN}\alpha_{\omega }^{*}\rho _{B}}{m_{\omega }^{*2}},  \label{v0}
\end{equation}
\noindent with the scalar and baryonic nuclear densities, $\rho _{s}$ and $\rho _{B}$, defined as usual as 
\begin{equation}
\rho _{s}=\left\langle \overline{\psi }\psi \right\rangle ,  \label{ros}
\end{equation}
\begin{equation}
\rho _{B}=\left\langle \psi ^{\dag }\psi \right\rangle .  \label{rob}
\end{equation}
\noindent For each value of the baryonic density, $\rho_s$ will be determined self-consistently by minimizing the energy density $\mathcal{E}$ (Eq.\ref{eps}) with respect to $\rho_s$.
\noindent In Eq.\ref{eqn}, $\Sigma _{0}$ represents the rearrangement term given by :

\begin{equation}
\label{sigma0}
\Sigma _{0}=-\frac{g_{\omega NN}^{2}\alpha_{\omega }^{*}\rho _{B}^{2}}{m_{\omega
}^{*}}\frac{\partial (\alpha_{\omega }^{*}/m_{\omega }^{*})}{\partial \rho _{B}}
-\rho _{s}\frac{\partial M_{N}^{*}}{\partial \rho _{B}}-\frac{g_{\sigma
NN}^{2}\alpha_{\sigma }^{*3}\rho _{s}^{2}}{m_{\sigma }^{*3}}\frac{\partial
(m_{\sigma }^{*}/\alpha_{\sigma }^{*})}{\partial \rho _{B}},  \label{rea}
\end{equation}
expression in which $\alpha_{\omega }^{*}$, $\alpha_{\sigma }^{*}$, $M_{N}^{*}$, $m_{\omega }^{*}$ and $m_{\sigma }^{*}$ are functions of $\rho _{B}$.
\noindent As in the standard Walecka model\cite{s.w}, Eq.\ref{eqn} is
equivalent to the Dirac equation for a nucleon of mass : 
\begin{equation}
m_{N}^{*}=M_{N}^{*}-g_{\sigma NN}\alpha_{\sigma }^{*}\phi _{0}.  \label{mns}
\end{equation}

Note that the vector potential $U_{0}$ is given by : 
\begin{equation}
U_{0}=g_{\omega NN}\alpha_{\omega }^{*}V_{0}-\Sigma _{0}.  \label{pov}
\end{equation}

As usual, the energy-density $\mathcal{ E}$, pressure $p$ and incompressibility
parameter $K$ can be obtained from the energy-momentum tensor, providing, for
 Fermi momentum $k_{F}$ : 
\begin{equation}
\mathcal{ E=}\frac{g_{\omega NN}^{2}\alpha_{\omega }^{*2}\rho _{B}^{2}}{2m_{\omega
}^{*2}}+\frac{g_{\sigma NN}^{2}\alpha_{\sigma }^{*2}\rho _{s}^{2}}{2m_{\sigma
}^{*2}}+\frac{\gamma }{16\pi ^{2}}\left[
k_{F}e_{F}^{*}(2k_{F}^{2}+m_{N}^{*2})-m_{N}^{*4}\ln \left( \frac{%
k_{F}+e_{F}^{*}}{m_{N}^{*}}\right) \right] ,  \label{eps}
\end{equation}

\begin{eqnarray}
p &=&\frac{g_{\omega NN}^{2}\alpha_{\omega }^{*2}\rho _{B}^{2}}{2m_{\omega }^{*2}}%
-\frac{g_{\sigma NN}^{2}\alpha_{\sigma }^{*2}\rho _{s}^{2}}{2m_{\sigma }^{*2}}%
-\rho _{B}\Sigma _{0}  \label{pres} \\
&&+\frac{\gamma }{48\pi ^{2}}\left[
k_{F}e_{F}^{*}(2k_{F}^{2}-3m_{N}^{*2})+3m_{N}^{*4}\ln \left( \frac{%
k_{F}+e_{F}^{*}}{m_{N}^{*}}\right) \right] ,  \nonumber
\end{eqnarray}
\begin{equation}
K=9\rho _{B}\frac{\partial ^{2}\mathcal{ E}}{\partial \rho _{B}^{2}},
\label{com}
\end{equation}

\noindent where $e_{F}^{*}=\sqrt{k_{F}^{2}+m_{N}^{*^{2}}}$ and $\gamma =4$
in symmetric nuclear matter. As already proved by Brown, Song, Min and Rho \cite{son}, the thermodynamical consistency holds, whatever the density dependence used in Eq.\ref{lag}, as long as the rearrangement terms (Eq.\ref{rea}) are not forgotten. 

\section{Quark dynamics in a Nambu Jona-Lasinio model}

\subsection{In medium quark mass} 

We consider the following chirally invariant two-flavor NJL Lagrangian\cite{njl} :

\begin{eqnarray}
\mathcal{ L}_{NJL} &=&\overline{q}\left[ i\gamma _{\mu }\partial ^{\mu }-m_{0}\right]
q+g_{1}\left[ (\overline{q}q)^{2}+(\overline{q}i\gamma _{5}{\bf \tau }%
q)^{2}\right] -g_{2}(\overline{q}\gamma _{\mu }q)^{2}  \label{njl} \\
&&+g_{3}\left[ (\overline{q}q)^{2}+(\overline{q}i\gamma _{5}{\bf \tau }%
q)^{2}\right] (\overline{q}\gamma _{\mu }q)^{2},  \nonumber
\end{eqnarray}

\noindent where $q$ denotes the quark field with two flavor ($N_{f}=2$) and
three color ($N_{c}=3$) degrees of freedom and $m_{0}$ is the diagonal
matrix of the current quark masses (here in the isospin symmetric case). The
second and third terms of Eq.\ref{njl}  represent local four-quark interactions while the later
one  is an eight-quark interaction. As explained in Section 4, the eight-quark term, while contributing very weakly to the energy, is essential here because its influence on the density dependences of the hadronic Lagrangian (Eq.\ref{lag}) enables to stabilize nuclear matter. In fact, this term provides an additional density dependence of the Lagrangian parameters which is important for the scalar polarizability of the nucleon. It is interesting to note that such a stabilizing effect has  already been reported by Bentz and Thomas\cite{b.t} in a quite different context, and by Mishustin et al. in a nucleonic NJL model \cite{mishu}. Other high-order interaction terms could be added to the Lagrangian \ref{njl}. The effect of some of them is actually under consideration but, in this exploratory work, only that appearing in \ref{njl} will be taken into account.

The Dirac equation for a quark in mean-field approximation  is
given by :

\begin{equation}
\left[ i\gamma _{\mu }\partial ^{\mu }-m_{0}-2 g_2\gamma_{0}\left\langle \overline{q} \gamma^0 q \right\rangle +2g_{1}\left\langle \overline{q}
q\right\rangle +2g_{3}\left\langle \overline{q}q\right\rangle \left\langle 
\overline{q}\gamma _{0}q\right\rangle ^{2}\right] q=0,  \label{dir}
\end{equation}

\noindent which defines a dynamical constituent-quark mass :

\begin{equation}
m=m_{0}-2g_{1}\left( 1+\frac{g_{3}N_{f}^{2}N_{c}^{2}\rho _{B}^{2}}{4g_{1}}
\right) \left\langle \overline{q}q\right\rangle =m_{0}-\widetilde{g}
_{1}\left\langle \overline{q}q\right\rangle ,  \label{gap}
\end{equation}

\noindent generated by a strong scalar interaction of the quark with the
QCD vacuum. In the gap equation (Eq.\ref{gap}), the quark condensate $\left\langle \overline{q}q\right\rangle $ can be written as :

\begin{equation}
\left\langle \overline{q}q\right\rangle =-i\int \frac{d^{4}k}{\left( 2\pi
\right) ^{4}}\texttt{Tr}S(k),  \label{qqbs}
\end{equation}
where here Tr denotes traces over color, flavor and spin. In Eq.\ref{qqbs}, $S(k)$ represents the in-medium quark propagator defined as :

\begin{equation}
S(k)=\frac{1}{\gamma _{\mu }k^{*\mu }-m+i\varepsilon }+i\pi \frac{\gamma
_{\mu }k^{*\mu }+m}{E^*_{k}}\delta \left( k_{0}-E_{k}\right) \theta \left(
k_{F}-\left| {\bf k}\right| \right) ,  \label{pro}
\end{equation}

\noindent where $k^*_{\mu} = k_{\mu} - 2(g_2-g_3\left\langle \overline{q}  q \right\rangle)\left\langle \overline{q} \gamma_{\mu} q \right\rangle $, $E^*_{k}=\sqrt{\mathbf{k}^{*2}+m^{2}}$, $E_k = E^*_k + 2(g_2-g_3\left\langle \overline{q}  q \right\rangle)\left\langle \overline{q} \gamma_0 q \right\rangle$ and $k_F$ is the quark Fermi momentum. It has to be noted that we are working in quark matter in the NJL model. The baryonic density is related to the total quark density by $ \rho_B = \frac{1}{3}\rho_q $.
\noindent The quark condensate is divergent due to the loop integrals
and requires an appropriate regularization procedure. As many authors\cite{b.m,bub}, we introduce a three-momentum cutoff $\Lambda $ which has the
least impact on medium parts of the regularized integrals, in particular at
zero temperature\cite{bub}. In fact, since the model is non renormalisable, the cut-off $\Lambda$ is just an additional parameter.

Thus, after the regularization procedure, the
quark condensate is given at each density by :

\begin{equation}
\left\langle \overline{q}q\right\rangle =-\frac{N_{f}N_{c}}{\pi ^{2}}%
\int_{k_{F}}^{\Lambda }\frac{mk^{2}dk}{E_{k}},  \label{qqb}
\end{equation}
self-consistent equation which is equivalent to a minimization of the energy density of quark matter with respect to the quark condensate.
\subsection{Mesons masses and couplings} 

As usual, the mesons  will be obtained by solving the Bethe-Salpeter
equation in the quark-antiquark channel.  Since this is the standard procedure, we will only list the results which are needed later on. First, we define the quark-antiquark polarization operator in the $M=\pi ,\sigma ,\omega $ channel by :

\begin{equation}
\Pi _{M}(q^{2})=-i\int \frac{d^{4}p}{\left( 2\pi \right) ^{4}}\texttt{Tr}
\left[ \Gamma _{M}iS(p+q/2)\Gamma _{M}iS(p-q/2)\right] ,  \label{pola}
\end{equation}
where the vertex $\Gamma _{M}=i\gamma _{5}\tau ^{k}$, $i \mathit{1}$ and $i\gamma
^{\mu }$ stand for respectively the pion, sigma and omega mesons. Note that,
in the vector channel, the Lorentz structure of the polarization operator is 
$\Pi _{\omega }^{\mu \upsilon }(q^{2})=\left( -g^{\mu \nu }+\frac{q^{\mu
}q^{\upsilon }}{q^{2}}\right) \Pi _{\omega }(q^{2})$. The in-medium meson
masses, $m_{M}^{*}$, and meson-quark-quark coupling constants, $g_{Mqq}^{*}$%
, are then determined by the pole structure of the $T$-matrix,
i.e. by the conditions:

\begin{equation}
1-K_{M}\Pi _{M}(q^{2}=m_{M}^{*2})=0,  \label{pol}
\end{equation}

\begin{equation}
g_{Mqq}^{*2}=\left[ \frac{d\Pi _{M}(q^{2})}{dq^{2}}\right]
_{q^{2}=m_{M}^{*2}}^{-1},  \label{res}
\end{equation}

\noindent where $K_{M}=\widetilde{g}_{1}$, $\widetilde{g}_{1}$ and $%
2g_{2}-2g_{3}\left\langle \overline{q}q\right\rangle ^{2}$ respectively for
the $\pi$, $\sigma$ and $\omega$ mesons. Note that for the determination of the
polarization operator (Eq.\ref{pola}), we have used the same regularization
procedure as for the quark condensate.

In the scalar channel, the polarization is given by: 

\begin{equation} 
\Pi_{\sigma}(q^2) = -\frac{\left\langle \overline{q}q\right\rangle}{m} + N_cN_f( q^2-4m^2) I(q^2),
\end{equation} 

\noindent with 

\begin{equation} 
 I(q^2) = \frac{1}{8\pi^2}\int_{4(\Lambda^2+m^2)}^{4(p_F^2+m^2)}\frac{1}{q^2-\kappa^2}\sqrt{1-\frac{4m^2}{\kappa^2}}d\kappa^2.
\end{equation} 

\noindent In the vector channel, one obtains: 

\begin{equation} 
\Pi_{\omega}(q^2) =-N_c N_f \frac{q^2}{12\pi^2}\int_{4(\Lambda^2+m^2)}^{4(p_F^2+m^2)}\frac{1}{q^2-\kappa^2}(1+\frac{2m^2}{\kappa^2})\sqrt{1-\frac{4m^2}{\kappa^2}}d\kappa^2.
\end{equation}

\noindent We also need the pion mass and decay constant to adjust the model parameters. In the pseudo-scalar channel, the polarization reads:  

\begin{equation} 
\Pi_{\pi}(q^2) = \frac{\left\langle \overline{q}q\right\rangle}{m} + N_cN_f q^2 I(q^2),
\end{equation} 

\noindent and the pion decay constant is : 

\begin{equation} 
 f_{\pi} = N_cN_fg_{\pi qq}mI(q^2=m_{\pi}^2).
\end{equation} 

\subsection{Density dependences for the hadronic model} 

We now have to define how the density dependences of the hadronic Lagrangian parameters (Eq. \ref{lag}) are deduced from quantities calculated in NJL model. First, for every value of the baryonic density, the $\omega$ and $\sigma$ meson masses $m^*_{\omega}$, $m^*_{\sigma}$ are taken directly from the NJL calculation (Eq. \ref{pol}) in which the value of the quark condensate has been obtained solving Eq.\ref{qqb}. Second, we have to determine $M^*_N$. Even if progress have been made (see for example \cite{b.t,pepin}), a realistic description of the nucleon in a NJL model is still an open problem.  Here, we have assumed that $M_{N}^{*}$ is directly related to the quark condensate,  with the same relation as found in finite-density QCD sum-rule calculations\cite{coh}. To leading order, which should be valid at densities below and around the saturation density of nuclear matter, one has then :

\begin{equation}
\frac{M_{N}^{*}}{M_{N}}= \frac{\left\langle \overline{q}q\right\rangle 
}{\left\langle \overline{q}q\right\rangle _{0}},  \label{sumr}
\end{equation}

\noindent where $\left\langle \overline{q}q\right\rangle$ is the value of
the  quark condensate at baryonic density $\rho_B$ obtained by solving Eq.\ref{qqb},  $\left\langle \overline{q}q\right\rangle _{0}$ is the same quantity in vacuum  and $M_N=939$ MeV is the free nucleon mass. 

It was predicted in \cite{br2} that the dependence of the vector-meson mass on the quark condensate should change around saturation from approximately $\left\langle \overline{q}q\right\rangle^{\frac{1}{2}} $ below saturation to $\left\langle \overline{q}q\right\rangle$ above. The dynamical processes leading to this change would probably also affect the nucleon mass at the same density. However, since in this work we are interested only in densities below and around saturation, we have assumed that Eq. \ref{sumr} remains valid at every density considered.
\noindent Third, in a simple $\mathrm{N_c}$ counting in a naive quark model, the quark-meson and nucleon-meson couplings in vacuum are proportional, as argued in \cite{b.m}. At low densities, it seems quite reasonable to assume that the same relation holds approximately. Such a proportionality holds also, for example, in the QMC model\cite{gui,sai}. The $\alpha_{\sigma }^{*}$ and  $\alpha_{\omega }^{*}$ functions entering in the hadronic Lagrangian would thus be taken as:
\begin{equation}
\alpha_{\sigma }^{*} = \frac{g_{\sigma qq}^{*}}{g_{\sigma qq}},  \label{rgs}
\end{equation}

\begin{equation}
\alpha_{\omega }^{*} = \frac{g_{\omega qq}^{*}}{g_{\omega qq}},  \label{rgw}
\end{equation}
\noindent where $g_{\sigma qq}^{*}$, $g_{\omega qq}^{*}$ and their values in vacuum  $g_{\sigma qq}$, $g_{\omega qq}$, are obtained from Eq.\ref{res} with the value of the quark condensate which minimizes the energy density of quark matter.

\section{Results and discussion}
\subsection{NJL model parameters} 
At the quark level, we have to fix five parameters for the NJL model : the
cutoff $\Lambda $, the bare quark mass $m_{0}$, and the coupling constants $
g_{1}$, $g_{2}$ and $g_{3}$. For a given value of the cutoff $\Lambda $, we need four constraints.
As usual, the pion mass $m_{\pi }=135$ MeV,  the pion decay constant $f_{\pi }=92.4$ MeV and the $\omega $ meson mass $m_{\omega }=782$ MeV in vacuum  are used. 
 Since the eight-quark term has been introduced in order to stabilize nuclear matter, at least one in-medium physical quantity should be constrained in the fitting procedure. We have chosen here to take into account the recent result obtained by the TAPS collaboration\cite{tap} for the $\omega $ meson mass in nuclei, $m_{\omega }^{*}(\rho_{B}=0.6\rho _{0})=722_{-4}^{+4}$ (stat)$_{-5}^{+35}$(syst) MeV (where $\rho _{0}$ is the saturation density of nuclear matter). 
Thus, the parameters $m_{0}$, $g_{1}$, $g_{2}$ and $g_{3}$  are obtained by fitting the pion and the $\omega $ meson properties in vacuum and the in-medium $\omega $ meson mass $m_{\omega }^{*}(\rho _{B}=0.6\rho _{0})=722$ MeV. We have verified that the results are qualitatively and quantitatively similar if we choose another constraint on the in-medium $\omega$ mass in accordance with the recent experimental data of Naruki et al.\cite{kek} on 12 GeV proton-nucleus reactions.
With these constraints, the cutoff $\Lambda$ cannot be unambiguously determined and, as often done, we have considered here several different values.\\

\subsection{Hadronic constraints and observables} 

On the other hand, at the hadronic level, $g_{\sigma NN}$ and $g_{\omega NN}$
have been determined  to reproduce the empirical saturation point of nuclear matter, i.e., $\mathcal{ E}/\rho_{B}-M_{N}=-16$ MeV and $p=0$ at $\rho _{B}=0.17$ fm$^{-3}$.

For each value of $\Lambda$ considered, in order to probe the description of nuclear matter obtained, we have calculated several physical quantities: the effective nucleon mass $m_{N}^{*}$, the incompressibility parameter $K$ and the slope of the real part of the energy dependence of the nucleon-nucleus optical potential  $U_{0}/M_{N}$ at saturation.
 For the compressibility of nuclear matter, the main information
comes from the energies of the monopolar modes over many nuclei. The commonly
accepted value $K=210\pm 30$ MeV, has been deduced by Blaizot\cite{bla}, in
the framework of non-relativistic Hartree-Fock RPA models with effective
interactions. In a relativistic framework, it has been shown\cite{ma} that
some effective Lagrangians leading to higher values of the incompressibility
 (close to $300$ MeV) could predict correctly the isoscalar giant
resonance energies in medium and heavy nuclei. Thus, we can estimate that a
realistic value of the incompressibility parameter should be in the interval: $K=250\pm 50$ MeV. Concerning the effective nucleon mass, Furnstahl, Rusnak
and Serot\cite{frs} have shown that for models without isoscalar tensor
coupling, there is a strong correlation between the value of the nucleon
effective mass at saturation density and the spin-orbit splitting in nuclei.
They found that, to accurately reproduce the empirical splittings, one
requires a value of $m_{N}^{*}/M_{N}$ between $0.58$ and $0.64$ at
saturation density. Finally, the slope of the energy dependence of
the real part of the nucleon-nucleus optical potential, which in a relativistic
mean-field approximation is $ U_{0}/M_{N} $, has been extracted\cite{f.s,s.j,dcm}
from experimental data\cite{ham} and has been found to be equal to $
0.30 $ in Ref.\cite{f.s,s.j} and $0.35$ in Ref.\cite{dcm}, up to $100$ MeV of
incident kinetic energy, at saturation. Thus, it is reasonable to expect that realistic values of $U_{0}/M_{N}$ should lie
between $0.25$ and $0.40$.

\subsection{Numerical results} 
Each value of $\Lambda$ corresponds to a value of the constituent quark mass in vacuum, and we have thus chosen to display the results as functions of $m$ instead of $\Lambda$. With the constraints chosen here at the quark level, the saturation point cannot be reproduced  for $m \leq 460$ MeV. We have plotted on Fig.1 the dimensionless nucleon effective mass $m_{N}^{*}/M_{N}$, the
incompressibility parameter $K$ and the slope of the energy dependence of the real part of the
nucleon-nucleus optical potential  $U_{0}/M_{N}$ at saturation  as 
functions of  $m$. 

\begin{figure}[htb]
	\centering
		\epsfig{file=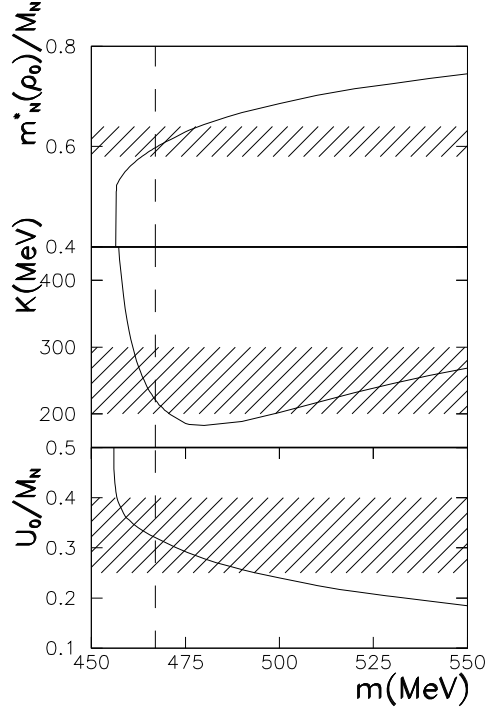,scale=0.5}
	\caption{Dimensionless effective nucleon mass $m_{N}^{*}/M_{N}$,
incompressibility parameter $K$ and slope of the energy dependence of the nucleon-nucleus
optical potential $U_{0}/M_{N}$ shown as functions of the constituent quark
mass $m$. The shaded areas correspond to the bounds on the empirical values
(see text) and the vertical dashed line indicates the central value $m=467$ MeV. }
	\label{fig1}
\end{figure}

The shaded areas correspond to the bounds on the empirical
values. As we can see, for $m\approx $ $465-470$ MeV ($\Lambda = 572 \pm 1$ MeV), the three physical quantities $m_{N}^{*}/M_{N}$, $K$ and $U_{0}/M_{N}$ are all in good agreement with the empirical values, and keep reasonable values at least up to $m \approx 500$ MeV.  

As an example, we have chosen to consider the central value $m=467$ MeV represented by the
 vertical dashed line in Fig.1. The parameters of the NJL model are then $\Lambda =572$
MeV, $m_{0}=5.6$ MeV, $g_{1}=8.22$ GeV$^{-2}$, $g_{2}=21.4$ GeV$^{-2}$ and $g_{3} = 1.08\  10^4$ GeV$^{-8}$. At the hadronic level, the values of the coupling constants are $g_{\sigma NN}=6.3$ and $g_{\omega NN}=13$. The saturation curve for symmetric nuclear matter is shown on Fig.2. 

\begin{figure}[htb]
	\begin{center}
		\epsfig{file=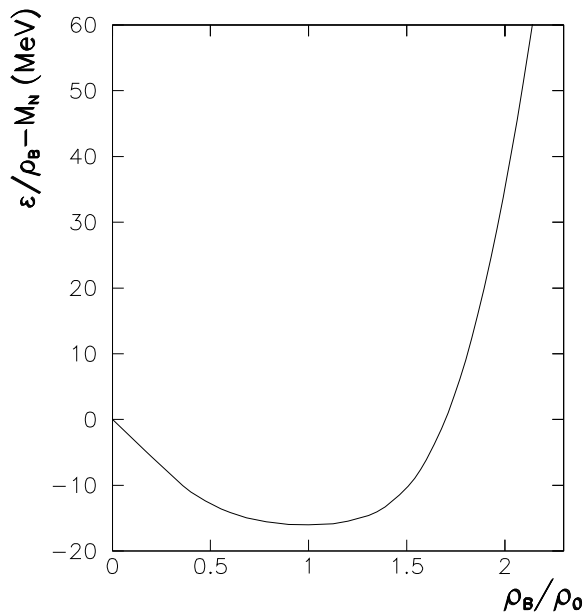,scale=0.5}
\end{center}		
		\caption{Saturation curve of nuclear matter obtained for $m=467$ MeV.}
\label{fig2}
\end{figure}

The values obtained for the effective nucleon mass, the incompressibility parameter and the slope of the optical potential at saturation are $m_{N}^{*}/M_{N}=0.6$, $K=221$ MeV, and  $U_{0}/M_{N}=0.32$. 
 Some remarks are in order.

First, the value of the $g_{3}$ parameter is such that $g_{3}N_{f}^{2}N_{c}^{2}\rho
_{B}^{2}/4g_{1}=0.02$ providing an eight-quark term which contributes only
for 2\% of the constituent quark mass at saturation (see Eq.\ref{gap}). However, this contribution is essential for stabilizing the system. With $g_{3}=0$, nuclear matter would saturate with an unrealistically weak effective nucleon mass at saturation density ($\approx $ 30\% of the free nucleon mass). The eight quark term tends to smooth the curvature of the quark condensate (and that of $M^*_N$ in this model) with respect to density. The sensitivity of the model to this term can be understood as follows. The saturation is resulting as a compensation between the scalar, vector and rearrangement term (Eq. \ref{sigma0}) contributions in pressure, and the rearrangement term is very sensitive to the derivatives of the masses and couplings with respect to density. The additional density dependence produces a variation of the $\Sigma_0$ term of about 40$\%$ at saturation. So, the difference between $g_3=0$ and $g_3\neq0$ lies principally in the modification of the derivatives of the masses and couplings rather than of these functions themselves. 

Second, at the quark
level, the model provides a quark condensate in the vacuum $\left\langle 
\overline{u}u\right\rangle ^{1/3}=-240$ MeV in good agreement with the
lattice calculations: $\left\langle \overline{u}u\right\rangle
^{1/3}=-(231\pm 4\pm 8\pm 6)$ MeV\cite{lat}. 

Third, the constituent quark mass in vacuum is relatively large. This value is appreciably larger than one third of the free nucleon mass, but since this quantity is not an observable, it should be model-dependent. Note that such large values are commonly considered in recent NJL calculations (see for example \cite{bub,oer}).  In particular, a large value has also been obtained\cite{oer} in a NJL model including meson loop corrections generated via a $1/N_{c}$
expansion of the self-energy in the next to leading order, with parameters
constrained by the experimental data on the electromagnetic form factor $F_{\pi} $ in the time-like region. Such a large mass doesn't prevent from obtaining a good description of the baryonic spectrum, for example using a quark-diquark model as done in ref \cite{buck,ishii}. Moreover, a 
large quark mass prevents the omega meson to be unstable against decay into a
quark-antiquark pair since $m_{\omega }^{*}$ is always lower than $2m$ for
each density.

Fourth, one can see on Fig. \ref{fig2} that the saturation curve is rather hard at densities above $\rho_0$. This can be due to the fact that $M^*_N$, taken here proportional to the quark condensate, is decreasing too fast at high densities. This could be compensated, for example, by introducing perturbatively higher order terms in the scalar and vector fields, as done in \cite{son}. 

\section{Conclusion}

We have investigated the  properties of nuclear matter in a
relativistic mean field model with density-dependent masses and couplings, similar to that used in \cite{son}, in which we have replaced the Brown and Rho scaling by a direct calculation of meson masses and couplings in a NJL quark model. The NJL model including four and eight quark interaction terms has been constrained to reproduce the recent TAPS result for the in medium $\omega$ meson mass in addition to the vacuum pion and $\omega$ meson properties. At the hadronic level, the two free parameters have been fixed to reproduce the empirical saturation point. 

At the quark level, the eight quark term is essential for obtaining realistic saturation properties. At saturation, the nucleon effective mass, the incompressibility parameter and the slope of the energy dependence of the nucleon-nucleus optical potential obtained are all in good agreement with the empirical data. 

Even if more work is needed in this direction and in a more fundamental one, this result is a very encouraging one since with only a few free-parameters, a realistic description of saturation properties of nuclear matter has been obtained from a hadronic Lagrangian constrained by a quark model which reproduces vacuum and in-medium meson properties.

\

\end{document}